\documentclass[AMA,STIX1COL,doublespace]{WileyNJD-v2}

\articletype{Article Type}%

\received{ NA}
\revised{ NA}
\accepted{NA }

\newcommand{\bs}[1]{{\bf{#1}}}
\newcommand{\bss}[1]{{\boldsymbol{#1}}}

\usepackage[figuresright]{rotating}
\usepackage{amsbsy, amsmath, amssymb, setspace}
\usepackage{multirow, color}
\usepackage{algorithm}
\usepackage{algpseudocode}

\begin{document}

\title{Analyzing Highly Correlated Chemical Toxicants Associated with Time to Pregnancy Using Discrete Survival Frailty Modeling Via Elastic Net}

\author[]{Abhisek Saha}
\author[]{Rajeshwari Sundaram *}
\authormark{SAHA AND SUNDARAM
}

\address[]{\orgdiv{{\it Eunice Kennedy Shriver} National Institute of Child Health and Human Development}, \orgname{National Institutes of Health}, \orgaddress{\state{Maryland}, \country{USA}}}



\corres{*Rajeshwari Sundaram,
	{\it Eunice Kennedy Shriver} National Institute of Child Health and Human Development,  National Institutes of Health, Bethesda, Maryland, USA \\ \email{sundaramr2@mail.nih.gov}}


\abstract[Summary]{
	Understanding the association between mixtures of environmental toxicants and time-to-pregnancy (TTP) is an important scientific question as sufficient evidence has emerged about the impact of individual toxicants on reproductive health and that individuals are exposed to a whole host of toxicants rather than an individual toxicant. Assessing mixtures of chemicals effects on TTP poses significant statistical challenges, namely (i) TTP being a discrete survival outcome, typically subject to left truncation and right censoring, (ii) chemical exposures being strongly correlated, (iii) accounting for some chemicals that bind to lipids, (iv) non-linear effects of some chemicals, and (v) high percentage concentration below the limit of detection (LOD) for some chemicals. We propose a discrete frailty modeling framework (named Discnet) that allows selection of correlated exposures while addressing the issues mentioned above.  Discnet is shown to have better and stable FN and FP rates compared to alternative methods in various simulation settings. We did a detailed analysis of the LIFE Study, pertaining to polychlorinated biphenyls and time-to-pregnancy and found that older females, female exposure to cotinine (smoking), DDT conferred a delay in getting pregnant, which was consistent across prior sensitivity analyses to account for LOD as well as non-linear associations. }

\keywords{Discrete survival; Frailty, Elastic net, Time-to-pregnancy, Chemical exposure, Limit of detection}


\maketitle


\section{Introduction}
\label{intro}
Synthetic environmental persistent chemicals, such as organochlorine pesticides, polychlorinated biphenyls and some metals have been reported to have endocrine-disrupting properties according to the Endocrine Society. \citep{diamanti2009endocrine,kortenkamp2008low} Growing evidence indicates exposures to endocrine-disrupting chemicals (EDCs) may adversely affect human health such as reproduction, metabolism to name a few. EDC exposure is widespread as the result of environmental distribution through natural processes and bioaccumulation long after halted industrial production. Furthermore, continual increases in the production of commonly used materials and goods containing a variety of EDCs that remain on the market, also contributes to environmental ubiquity. \cite{bergman2013world} 
As a prominent example, the global production of plastics, which contains numerous EDCs, has continually increased for the past five decades from 1.5 million tons to more than 300 million tons \citep{geyer2017production} confirming the ubiquitous presence of multiple EDCs in the contemporary environment, and therefore, increases risks of adverse reproductive health.  Data from human biomonitoring studies suggests women of childbearing age are often exposed to multiple EDCs simultaneously. \citep{mitro2015cumulative, woodruff2011environmental} Thus, this necessitates detailed study quantifying the risk of adverse reproductive outcomes in the context of EDC mixtures. Motivated by these issues, we are interested in studying the association of mixtures of EDCs with human fecundity, the biologic capacity to reproduce by men and women. \citep{gini1924premieres} 

Assessing EDC mixtures association with human fecundity presents multiple statistical challenges. Human fecundity is quantitatively assessed through time-to-pregnancy (TTP), the number of menstrual cycles needed to get pregnant by non-contraceptive couples.  TTP is a discrete survival time, subject to left truncation as well as right censoring due to various enrollment criteria of pregnancy studies. Additionally, it is well known that human fecundity has significant unmeasured heterogeneity. \cite{ecochard2006heterogeneity} This necessitates including frailty in the model for TTP.  While EDCs are highly correlated within an individual, as well as with the partners’ exposure levels, they further present quite a many challenges when it comes to modeling risk, such as (i) high concentration of values below limits of detection (LOD) in ascertaining the EDC levels: - machines reported values may not be reliable, (ii) lipophilic nature of some EDCs, i.e, they bind to the lipids in the blood: - high amount of lipid will induce high amount of chemicals, (iii) suspected non-linear effects of some EDCs and (iv) missingness in data. As modern experiments are capable of collecting large number of EDCs, modeling risk in the context of studying EDC mixtures requires a variable selection strategy which can handle the nuances of such data. 

 To bridge this gap, we propose a variable selection approach, called \texttt{discnet} to assess the mixtures of EDCs on TTP, which allows us to identify the ``important drivers” of the mixtures of EDCs on human fecundity. \citep{lazarevic2019statistical} To the best of our knowledge such analysis was never performed before due to absence of variable selection methods addressing the mentioned issues. In this context our proposed method is novel, and leads to novel findings, that are consistent to prior research done in restricted settings.  
 
 We first note that such a problem can be converted to a penalization regression framework with modified generalized linear mixed model  prior to adjusting for LOD and lipid binding phenomena.  
Although there have been significant literature on variable selection using penalized regression framework spanning over few decades (in linear model,  \citep{tibs1996, zou2005enet}  in generalized linear model, \citep{park2007l1, rosset2004following, shevade2003simple} in continuous cox proportional hazard model \citep{goeman2010L1, tibs1997cox}),
relatively, far less focus was given to variable selection in the context of generalized linear mixed models (GLMM). In this context Groll and Tutz's work is noteworthy as they developed L1-penalty based variable selection for GLMMs. \citep{groll2014variable} They later adapted it for variable selection for discrete frailty model, \citep{groll2017variable} which we here refer to as \texttt{glmmlasso}. L1-based method usually performs poorly in presence of highly correlated variables, \citep{zou2005enet} and they do not produce stable paths. \citep{park2007l1}  In our proposed framework \texttt{discnet}, we propose fast variable selection method based on elastic net penalty that can incorporate left truncation while addressing the concerns such as LOD and lipid treatment by incorporating cutting-edge procedures on the covariates. Importance of elastic net penalty to deal with high dimensional correlated covariates in discrete frailty modeling is highlighted through  extensive simulations showcasing its effectiveness  compared to \texttt{glmmlasso} as it produces comparable FN rates while improving upon FP rates and stability of the results. We have extended our methodology to grouped selection to include categorical variables, and have shown the performance remains equally good.  Further, while implementing for real data-analysis we developed multiple versions of the approach to incorporate multiple cutting edge methods to deal with the same issues, such as high below LOD values or chemicals binding with lipid etc.  Our methodology also differs from that of \texttt{glmmlasso} in how penalty parameters are tuned.  We have the key R-function, \texttt{Discnet.R} with necessary library support of auxiliary R-functions, which can be supplied on request. 

In Section 2, we present our data structure and proposed penalized framework. In Section 3, we present necessary estimation framework that discusses computational schemes and, finally extend it to group selection.  In Section 4 we present detailed finite sample investigation of our proposed approach through extensive simulations, and evaluate its strengths compared to the competing method. In Section 4 we have thoroughly investigated our proposed variable selection approach in the context of mixtures of EDCs, including Polychlorinated Biphenyls, Cadmium, Mercury and Lead from both partners and their time-to-pregnancy from the LIFE Study. \citep{buck2011designing} We have systematically addressed the issues of LOD, lipid adjustment to account for lipophilic EDCs and issues for non-linearity, as well as account for missingness through multiple imputation. Finally, in Section 6, we summarize our novel contributions, and key findings. 

\section{Method}
\label{DHazard}

Let $T$  be the discrete time to event (e.g., number of menstrual cycles needed to get pregnant), taking values in $\{ 1, \cdots, k\}$. let $\bs{x} $ be the vector of underlying covariates.  In discrete survival analysis, one commonly models the discrete hazard function, $\lambda(t|\bs{x})$, and estimates the survivor function S(t),
\begin{eqnarray*}
	\lambda(t| \bs{x}) &=& P(T =t | T \ge t, \bs{x}),  \label{haz.1}\\ 
	S( t | \bs{x} ) &=&  P(T \ge t| \bs{x}) = \prod_{i=1}^{t-1} (1- \lambda(t| \bs{x})), \quad  t= 1, \cdots k. \label{haz.2} 
\end{eqnarray*}
We consider here the log odds model for the hazard, which is the Cox model for the discrete survival \citep{cox1972regression} as given below,

\begin{eqnarray}
	\lambda(t| \bs{x}) = h(\gamma_{t} + \bs{x}^T \bss{\beta}), \quad h=  g^{-1} , \,\, g= \text{logit link function}   
	\label{Glm}
\end{eqnarray}
where $\gamma_{t}$ is unspecified baseline hazard,  and $\bss{\beta}$ is the vector of coefficients corresponding to  $\bs{x}$. We further assume $T$ is subject to both right censoring  and left truncation. Let $L$ and $C$ be the non-informative left truncation random variable and non-informative right random censoring variables  \citep{kalbfleisch2011statistical}. Note, in presence of left truncation $T$ is observed only if $T \ge L$. Due to right censoring, $min(T, C)$ is only observed. We further assume that $(L,C)$ is independent of $T$ and that $P(L \leq C)=1$. Let  $\delta$ be the censoring  indicator, $\delta = I\{T \leq C\}$ and $\widetilde{T}= min\{T,C\}$ be the observed time to event variable. Thus, $\delta=1$ implies  $L \le T \le C$ while $\delta=0$ implies   $ T > C$.  When $T < L$ the sample is  not observed. Let the $i^{th}$ sample be represented by the triplet $(t_i,l_i,\delta_i)$ for $(\widetilde{T}, L, \delta)$. 
Likelihood for a single observation $i$ can be written using (\ref{Glm}) as follows, 
\begin{eqnarray}
	L( \bss{\beta}, \bss{\gamma}_0 |\bs{y}_i)&=& \prod_{s=l_i}^{t_i} \lambda(s| \bs{x}_i)^{y_{is}} (1- \lambda(s| \bs{x}_i))^{(1-y_{is})}, \label{augModel}    
\end{eqnarray}
where $y_{is}$'s are augmented pseudo-observations such that  $\{ t_i, \delta_i=1, l_i \} \Rightarrow \{ y_{is} = 0; s=l_i , \cdots , t_i-1 ; y_{it_i}= 1 \}$ and $ \{ t_i, \delta_i=0,l_i \} \Rightarrow  \{ y_{is} = 0; s=l_i ,\cdots , t_i-1; y_{it_i}= 0 \}$. Thus, (\ref{augModel}) mimics a likelihood of binary logistic model with $t_i-l_i+1$ many observations.
\vspace*{-0.65 cm}
\subsection{Modeling hazard in presence of frailty}
\label{s:frailty}
In fecundity studies frailty is often introduced in modeling discrete survival variable such as time-to-pregnancy (TTP) to account for significant unmeasured heterogeneity at subject level. \citep{ecochard2006heterogeneity}   
Representing frailty for $i^{th}$ subject by $q$ random effects, say $\bs{b}_i$, where  $\bs{b}_i \sim N(\bss{0}, \bs{Q})$, discrete hazard, $\lambda(s| \bs{x}_i)$ from  (\ref{augModel}) can be given as follows,

\begin{eqnarray}
	\lambda(s| \bs{x}_{i}, \beta_0,\bss{\beta}, \bs{b}_i , \gamma_{s}) = h(\eta_{is})= h(\gamma_{s} + \beta_0+ \bs{x}_{i}^T \bss{\beta}+ \bs{z}_{is}^T \bs{b}_i )  \label{frailty} \\
	\eta_{is}= \gamma_{s}+\beta_0+ \bs{x}_{i}^T \bss{\beta}+ \bs{z}_{is}^T \bs{b}_i , s= 1, \cdots, t_i, i=1, \cdots, n  \nonumber 
\end{eqnarray}

where $\bs{z}_{is}$ is the set of possibly time-varying covariates with random effects. In general, one can have $\bs{x}_{i}$'s vary over time. Let us define $\bs{b}^T = (\bs{b}_1^T,\cdots, \bs{b}_n^T)$. Let $\bs{A}_{is}$ be a binary vector of length $t_{max}$ indicating position $s$,  and $\bs{A^T} , \bs{X^T}, \bs{Z^T}$ be the corresponding matrices by stacking rows one over the other.  Thus, $\bs{b} \sim \bs{G(b)} = N(0, \bs{Q}_b= I_n \bigotimes \bs{Q} ) $.

\subsection{Variable selection with correlated covariates}
\label{varsel}
Motivated by the fecundity study that measures environmental toxicants  in blood which are often highly correlated  as discussed in Section \ref{intro}, we assume that covariates may be potentially highly correlated. Coupled with the fact that the number of covariates p is moderately large compared to sample size, correlated covariates pose an extra challenge in estimating the regression coefficients as standard methods would fail to produce stable and efficient estimates. This necessitates a variable selection strategy. We propose a penalized regression framework with Elastic net penalty as, unlike Lasso, it gives equal weightages to all correlated variables and produces stable paths. \citep{zou2005enet,park2007l1} To obtain robust estimate on baseline hazard, we apply a ridge penalty  on $\bss{\gamma}$. Let $l( \beta_0,\bss{\beta},\bss{\gamma},\bs{Q} |\bs{y})$ be log of marginal likelihood (or aka integrated) where $\bs{y}'= [\bs{y}_1',\cdots,\bs{y}_n']$. The penalized likelihood can be given using (\ref{frailty}) as follows.       
\begin{eqnarray}
	l_{pen}(\beta_0,\bss{\beta}, \bss{\gamma},\bs{Q}| \bs{y}) &=&  l( \beta_0,\bss{\beta},\bss{\gamma},\bs{Q} |\bs{y}) - \nu(\alpha\sum_{j=1}^p|\beta_j| + (1 - \alpha) \sum_{j=1}^p\beta_j^2/2) \nonumber 
	- \nu_s \sum_{m=1}^{t_{max}}\gamma_{m}^2/2 , \nonumber \\
	&&\nu, \nu_s > 0, \quad  0 < \alpha < 1. \label{pen.rand}
\end{eqnarray}
where ($\nu$ and $\alpha$) are penalty parameters for Elastic net whereas   $\nu_s$ is the ridge penalty for the baseline parameters.

\vspace*{-0.5 cm}
\section{Estimation}
\label{s:Est}
Estimating parameters require special attention due to a) presence of random effects as the integrated likelihood does not admit a closed analytical form, b) non-differentiability of L1 penalty in Elastic net. Let $\bss{\theta}$ denote the set of all parameters except frailty parameters, $\bs{Q}$. 
\subsection{Maximum Likelihood (ML) Estimation}
Given a set of values for the tuning parameters, $\nu, \alpha$ and  $\nu_s$, we propose to find ML estimates by co-ordinate wise gradient  ascent approach for $(\bss{\theta},\bs{Q})$. We implement a (modified) gradient ascent approach \citep{goeman2010L1} for $\bss{\theta}$ fixing co-ordinates of $\bs{Q}$  whereas we block-update  $\bs{Q}$ using EM-based method \citep{groll2014variable} fixing co-ordinates of $\bss{\theta}$. We iterate  between these updates until convergence. For ML estimation it is sufficient to work with the following approximated penalized  likelihood with $qn$ many new augmented parameters $\bs{b}$, that can be obtained from (\ref{pen.rand}) by applying Laplace method on the integrated unpenalized likelihood  \citep{bres1993} part.   
\vspace{-0.5cm}
\begin{eqnarray}
	l_{pen}^{app}(\beta_0,\bss{\beta}, \bss{\gamma},\bs{b},\bs{Q} | \bs{y}) &=&  \log
	f(\bs{y}| \beta_0,\bss{\beta}, \bs{b},\bss{\gamma})-1/2\bs{b}^T\bs{Q}_b^{-1}\bs{b} 
	-\nu(\alpha\sum_{j=1}^p|\beta_j| + (1 - \alpha) \sum_{j=1}^p\beta_j^2/2)  
	- \nu_s \sum_{m=1}^{t_{max}}\gamma_{m}^2/2 , \label{penEq}
\end{eqnarray}
Thus, we define $\bss{\theta} = \{ \beta_0, \bss{\beta},\bss{\gamma},\bs{b}\}$. We have summarized the optimization algorithm for \texttt{Discnet}, in detail for logit link in Algorithm \ref{Myalgo} of Appendix B. The algorithm implements a path following mechanism on the grid of tuning parameters to find parameter estimates, and eventually, refits the model with the selected variables by optimizing (\ref{penEq}) with elastic net penalty removed for post-selection inference.

\subsection{Tuning and faster convergence} 
The parameters, $\nu, \nu_s$ and $\alpha$ require tuning together.  We first estimate $\nu_{perm}(\alpha, \nu_s)$ using permutation selection \citep{sabourin2015permutation} for a fixed pair of $(\alpha, \nu_s)$, where each such pair can be selected from a rectangular grid. The optimal combination can be chosen by minimizing the overall \emph{BIC($\nu_s,\alpha,\nu_{perm}(\nu_s,\alpha$))} or prediction error of cross-validated samples. For simplicity, one can use a value $\nu_s=100$ as recommended in  Groll et al\citep{groll2017variable} for real data scenarios. Once the variable selection is done, for post-selection inference the model with the selected variables  is fitted using Fisher scoring (while still penalizing the baseline with ridge penalty; only elastic net penalty is dropped at this stage).  In Algorithm \ref{Myalgo}, $\bss{\theta}$ update step (See 4. of \textbf{Step 2}) allows for a  Fisher scoring step that can help speed up convergence, but is only applicable when optimum step size is within maximum allowed step-size, and the chain has gotten into a region where the gradient is continuous. Thus, Fisher scoring fails when the chain moves around the region of discontinuity. Under these circumstances, it is recommended to let the gradient ascent run initially long enough until it lands into a region of continuous gradient, after which it can be replaced with iterated weighted least square Fisher's scoring.

\subsection{Adjustments for group selection with elastic net}
We next present how our proposed methods can be adapted to group selection.  We modify Meier et all (2008) \cite{meier2008group}'s approach to include a group specific $L_2$ norm penalty $||.||_2$  after we club  coefficients $\bss{\beta}$ into groups of variables. Let $\bss{\beta}_g$ be a sub-vector of $\bss{\beta}$ that corresponds to $g^{th}$ group. Let $m_g$ be the length of $\bss{\beta}_g.$  Then the penalized likelihood can be written as follows,  
\begin{eqnarray}
	l_{pen}(\beta_0,\bss{\beta}, \bss{\gamma},\bs{Q}| \bs{y}) &=&  l(\beta_0,\bss{\beta},\bss{\gamma}, \bs{Q} |\bs{y}) - \nu\sum_{g=1}^G (\sqrt{m_g}||\alpha\bss{\beta}_g||_2
	+ (1 - \alpha) \bss{\beta}_g'\bss{\beta}_g/2)
	\nonumber\\
	&&
	- \nu_s \sum_{m=1}^{t_{max}}\gamma_{0m}^2/2,\quad \bss{\beta}^T=(\bss{\beta}_1^T,\cdots,\bss{\beta}_G^T);\,\,  \nu, \nu_s > 0, \quad  0 < \alpha < 1 \nonumber \label{pen}
\end{eqnarray}
Note when $g=1$, $||.||_2$ becomes $L_1$ penalty. The estimation requires Algorithm \ref{Myalgo} to modify $\bss{\beta}$ update step. The changes are listed in Algorithm 2 as given in the Web Appendix B.  

\section{Simulation Study }
\label{simul}
We evaluate the effectiveness of our method called \texttt{Discnet}, in two different ways, \emph{Scenario I} and \emph{Scenario II} . In \emph{Scenario I}, we compare its performance with \texttt{glmmlasso} ( referred to as \texttt{glmmlasso\_discrete} in Groll et al (2017) \citep{groll2017variable})  in the context of frailty modeling for discrete survival analysis. Since our primary focus is to compare methods that deal with random effects models, there are not many options available to the best of our knowledge.  Most of the other approaches, that allow variable selection, do not allow us to incorporate ridge penalty on baseline directly. Groll et al (2017) compared \texttt{glmmlasso}  with approaches that allow random effects, such as \texttt{glmer} \citep{bates2014fitting},  \texttt{gamm4} \citep{wood2017package} in generalized linear mixed model set-ups adapting to stepwise variable selections.  They also noted that in absence of random effects, \texttt{glmmlasso} performs at par with popular approaches, such as \texttt{glmnet} \citep{friedman2010regularization}, \texttt{penalized} \citep{goeman2010L1}.   Our proposed method  implements elastic net penalty, which is critical for correlated variables, while incorporating left truncated observations, as well as more efficient  permutation-based  tuning on $\nu$. In \emph{Scenario II}, we evaluate effectiveness of  group selection across scenarios, when multiple groups are present along with correlated variables, and with left truncated right censored discrete survival time. 

\subsection{ Scenario I: Impact of left truncation and random right censoring on performance}
To be relevant for survival analysis as in public health studies, we work with fixed set of  $p=150$ many covariates.  We consider 12 cases based on three design parameters, namely, 1. $n$=sample size with two levels (150 ($n=p$), 250 ($n>p$)), 2. Cn= censoring proportion with three levels (Low=20\%, Med=35\%, and High=50\%) and 3. Tr = truncation indicator with two levels (0=No truncation, 1=40\% left truncation). We further assume only a few, $p^*=5$, many covariates are associated with discrete survival time $T$ such that $T$ takes values in \{1,\ldots, $T_{max}$=10\}. Thus, $T$ is simulated  from the following frailty model: \[\lambda_{i,t}= g(\gamma_{t} + \sum_{i=1}^{p}\beta_i* x_i+ r_i);\,  r_i \sim N(0,\sigma^2=1), g= \rm{logistic}(.) \] in which the vector of non-zero regression coefficients, $\bss{\beta}^{*}=(\beta_1,...,\beta_5)'=(-4,-4,-4,8,8)'$, and $\beta_i=0, p^{*} < i \le p$.  The time-varying  baseline hazard parameters, $\bss{\gamma}$ is assigned $(-9.00, -7.00, -4.97, -2.82,  0.34, 1.39,  2.35,  4.27,  6.18,  8.11)'$.       $\gamma_{t}$ is chosen an increasing function of $t$ such that there is good representation of various instances of survival time, $T$, at all levels of censoring. In presence of left truncation (Tr=1), left truncation variable ,$L$,  is simulated from \emph{Mult(0.6,0.2,0.2)} with labels 1,2,3 (with 1 indicating no truncation), motivated by real data as discussed in Section 5. Thus, for each choice of the triplet (n, Cn , Tr), 1000 data replicates  are generated from the model. Two approaches, namely \texttt{Discnet} and \texttt{glmmlasso} are compared based on average performance over 1000 data-sets on three popular measurement metrics, namely average number of False Negatives (FP), False Positives (FP) and median squared error of estimator for coefficient vector $\bss{\beta}$ (Med\_SE). One can obtain more commonly used quantities, such as False Positive Rate (FPR) and False Negative rate(FNR) with FP and FN,  divided by $p^*$ and $p-p^*$ respectively. While implementing $\texttt{glmmlasso}$, optimal model is chosen based on {\it{BIC}} criterion.\cite{groll2017variable}  In case of implementation of \texttt{Discnet},  solution path is first constructed over ($\nu, \nu_s, \alpha$).  while $\alpha$ and $\nu_s$ are tuned over the grid $ [1,0.95,0.8,0.7,0.6,0.5] \times [15,25,50,100]$ based on ${\it{BIC}}$, lasso parameter, $\nu$,  is tuned following permutation approach. \citep{sabourin2015permutation} 
We have implemented this method under two settings, 1) with independent covariates, and 2) with correlated covariates. Under setting 1) we have found that \texttt{discnet} was as effective as that of \texttt{glmmlasso} with improved performance in terms of FP rates and stability of results on some occasions. More details are given in Web Appendix C. 

We elaborate the setting 2) here in which we generate correlated variables $x_i's$ in the following manner. We first generate independent $z_i's$ from $U(0,1)$. We define $x_1=z_1,x_2= \theta_1 x_1 + z_2,x_3 = \theta_1 x_1 + z_3$. This implies $cor(x_1, x_2)=cor(x_1, x_3)= \frac{\theta_1}{\sqrt{1+\theta_1^2}}=\rho_1$  (therefore, $\theta_1=\frac{\rho_1}{\sqrt{1- \rho_1^2}}$) and $cor(x_2, x_3)= \rho_1^2$. If we choose $\rho_1=.9$, it corresponds to $\beta_1 \approx 2$. Continuing like this with $x_1$, one can define a block of variables (excluding $x_1$) with each pair having a fixed correlation $\rho_1^2$ (compound symmetry (CS)) leaving $x_1$. Thus, to define a CS with 2 blocks  of size 3 each, and between-block independence, one can use $x_i=\beta_1 z_{p+1} + z_i , i=1,...,3$, and  $x_i=\beta_2 z_{p+2} + z_i , i=4,5,6$. We make a note that although $z_i'$ are uniform $x_i's$ behave more like triangular distribution ( $U(0,1)+U(0,1) = Triangular(0,2)$)
In what follows, we have assigned 2-block correlation structure, each of size 3,  with $\rho_1=0.7$ and $\rho_2=.4$  in them respectively, while the rest of $p-6$ variables are generated from $U(0,1)$.  Out of these first 6 covariates only first five are associated with the outcome variable as mentioned earlier. Performance metrics are reported in Table \ref{tab2}

\begin{table}[h]
	\centering
	\caption{FP, FN and MSE$_{\bss{\beta}}$ (Med\_SE) based on 1000 replicates with correlated covariates ($R=0.7J_3+0.3I_3\bigodot 0.4J_2+0.6I_2$), for \texttt{Discnet} and \texttt{glmmlasso} respectively. p=150. \texttt{Discnet} outperforms \texttt{glmmlasso} consistently in terms of FP rates while FN rates remain comparable across three censoring levels; FN rates drop with the increase of n} \label{tab2}
	
	\begin{tabular}{lcccccccc}		
			\toprule
		\multicolumn{3}{c}{Cases} & \multicolumn{3}{c}{Discnet} & \multicolumn{3}{c}{glmmlasso} \\
		n     & Cn   & Tr  & FN    & FP    & Med\_SE  & FN     & FP    & Med\_SE  \\
		\midrule     
		150 & 0.21 & 0.34 & 0.00 (0.05) & 0.01 (0.11) & 100.55 (3.63) &0.00 (0.06) & 0.31 (0.65) & 100.61 (3.70) \\ 
		250 & 0.21 & 0.34 & 0.00 (0.00) & 0.01 (0.11) & 95.42 (2.72) &0.00 (0.03) & 0.50 (0.95) & 95.45 (2.80) \\ 
		150 & 0.36 & 0.33 & 0.02 (0.15) & 0.01 (0.09) & 98.31 (4.53) &0.02 (0.17) & 0.29 (0.58) & 98.55 (4.69) \\ 
		250 & 0.36 & 0.32 & 0.00 (0.00) & 0.01 (0.12) & 93.83 (3.34) &0.00 (0.00) & 0.47 (0.88) & 93.92 (3.40) \\ 
		150 & 0.50 & 0.30 & 0.09 (0.29) & 0.03 (0.19) & 93.24 (5.76) &0.07 (0.28) & 0.30 (0.57) & 93.59 (6.08) \\ 
		250 & 0.50 & 0.30 & 0.00 (0.06) & 0.02 (0.15) & 89.61 (4.41) &0.00 (0.06) & 0.46 (0.81) & 89.73 (4.51) \\ 
		\hline
		150 & 0.22 & \multirow{6}{*}{0.00} & 0.00 (0.07) & 0.00 (0.06) & 101.66 (3.41) &0.01 (0.08) & 0.26 (0.58) & 101.76 (3.47) \\ 
		250 & 0.22 &  & 0.00 (0.00) & 0.00 (0.07) & 95.96 (2.55) &0.00 (0.04) & 0.47 (0.91) & 95.96 (2.63) \\ 
		150 & 0.38 &  & 0.03 (0.16) & 0.01 (0.09) & 98.87 (4.45) &0.02 (0.15) & 0.30 (0.57) & 99.08 (4.58) \\ 
		250 & 0.38 &  & 0.00 (0.00) & 0.01 (0.12) & 94.31 (3.29) &0.00 (0.00) & 0.44 (0.85) & 94.32 (3.33) \\ 
		150 & 0.53 &  & 0.09 (0.28) & 0.03 (0.16) & 92.91 (6.18) &0.06 (0.28) & 0.28 (0.56) & 93.21 (6.44) \\ 
		250 & 0.53 & & 0.00 (0.07) & 0.02 (0.14) & 89.43 (4.42) &0.01 (0.08) & 0.46 (0.81) & 89.65 (4.47) \\ 
\bottomrule
\end{tabular}
\end{table}

We again observe a similar pattern as in case 1) as our proposed method consistently remains as effective as the other, specially improving upon in terms of FP rates while remaining comparable with respect to other FN rates and MED\_SE. From  Table \ref{tab2}, it is evident that the proposed methods perform very well even in presence of truncation.

\subsection{ Scenario II: Performance in group selection while other correlated covariates are present}

In this scenario our objective is to see how grouped elastic-net version performs across various scenario's. We borrow $\bss{\gamma}, \bss{\beta^{*}}$, Cn and  Tr  from earlier set-up as they are. Although we kept the same correlation structure of 5 un-grouped variables we added 7 variables, constituting two groups with first 4 and last 3 respectively of them, which are associated with the outcome. We consider three circumstances: 1) \texttt{cont}:  both of the groups continuous in nature, 2) \texttt{cat}: both of them representing two categorical variables, and 3) \texttt{mixed}: the first group represents categorical one while the next one is continuous in nature.  Thus, if first one is a categorical variable, it has 5 levels with the 5$^{th}$ one being treated as reference class. As a result, we consider the following vectors of regression coefficients corresponding to two groups as follows,     $\bss{\beta}_1^{*}= (7, -5,	7,-4)', \bss{\beta}_2^{*}= (5, -8,	3)'$. To measure performance  accuracy, we first define FP and FN as before based on all variables including  7 new sub-group variables as variables in their own right. We further define a non-group False Negative (NG\_FN)  if  a non-group variable with non-zero coefficient is mis-classified as 0. We further compute group capture percentage for each of them based on 1000 replicates. All the metrics are reported in Table \ref{tab3}.

\begin{table}[h]
	\centering
	\caption{FP, FN, Non-group FN (NG\_FN), MSE$_{\bss{\beta}}$, Percentage of GRP\_1 captured (GRP1) and GRP\_2, captured (GRP2) based on 1000 replicates with other correlated co-variates ($R=0.7J_3+0.3I_3\bigodot 0.4J_2+0.6I_2$). p=150. } \label{tab3}
	\begin{tabular}{lcccccccc}
		\toprule
		Type & n & Cn & FN & FP & NG\_FN  & Med\_SE & GRP1 & GRP2 \\\hline
		\multirow{6}{*}{cont}  & 150 & 0.20 & 0.90 (2.24) & 0.05 (0.23) & 0.81 (0.95) & 231.08 (38.85) & 88.90\% & 86.50\% \\ 
		& 250 & 0.20 & 0.00 (0.03) & 0.00 (0.03) & 0.40 (0.66) & 218.75 (7.19) & 100.00\% & 100.00\% \\ 
		& 150 & 0.36 & 3.11 (3.42) & 0.24 (0.49) & 0.68 (1.02) & 239.50 (62.06) & 59.40\% & 58.40\% \\ 
		& 250 & 0.36 & 0.00 (0.05) & 0.00 (0.05) & 0.60 (0.81) & 206.86 (10.35) & 100.00\% & 100.00\% \\ 
		& 150 & 0.49 & 4.62 (3.33) & 0.43 (0.65) & 0.56 (0.95) & 321.81 (78.29) & 41.60\% & 38.10\% \\ 
		& 250 & 0.49 & 0.12 (0.72) & 0.01 (0.11) & 0.88 (1.05) & 188.97 (24.79) & 99.10\% & 97.50\% \\ 
		\hline
		\multirow{6}{*}{cat}   & 150 & 0.22 & 0.09 (0.29) & 0.09 (0.29) & 0.67 (0.79) & 230.19 (8.97) & 100.00\% & 100.00\% \\ 
		& 250 & 0.22 & 0.00 (0.03) & 0.00 (0.03) & 0.39 (0.67) & 219.78 (6.17) & 100.00\% & 100.00\% \\ 
		& 150 & 0.33 & 0.17 (0.47) & 0.16 (0.38) & 0.76 (0.87) & 225.46 (12.52) & 99.80\% & 99.90\% \\ 
		& 250 & 0.33 & 0.01 (0.08) & 0.01 (0.08) & 0.47 (0.75) & 214.92 (7.52) & 100.00\% & 100.00\% \\ 
		& 150 & 0.49 & 0.52 (1.28) & 0.37 (0.62) & 0.86 (1.11) & 206.45 (29.86) & 97.50\% & 98.20\% \\ 
		& 250 & 0.49 & 0.04 (0.19) & 0.04 (0.19) & 0.58 (0.80) & 196.94 (11.97) & 100.00\% & 100.00\% \\ 
		\hline 
		\multirow{6}{*}{mixed}  & 150 & 0.21 & 0.27 (0.88) & 0.02 (0.14) & 0.75 (0.93) & 232.03 (18.28) & 100.00\% & 91.50\% \\ 
		& 250 & 0.21 & 0.00 (0.00) & 0.00 (0.00) & 0.42 (0.70) & 220.78 (6.47) & 100.00\% & 100.00\% \\ 
		& 150 & 0.37 & 1.17 (1.57) & 0.12 (0.34) & 0.61 (0.82) & 230.12 (31.27) & 99.90\% & 65.40\% \\ 
		& 250 & 0.37 & 0.01 (0.17) & 0.00 (0.06) & 0.62 (0.84) & 213.31 (9.09) & 100.00\% & 99.80\% \\ 
		& 150 & 0.51 & 2.42 (1.86) & 0.37 (0.59) & 0.51 (0.78) & 262.87 (1573.51) & 98.00\% & 34.20\% \\ 
		& 250 & 0.51 & 0.09 (0.53) & 0.01 (0.10) & 0.78 (0.94) & 197.42 (16.12) & 100.00\% & 97.20\% \\  
\bottomrule
\end{tabular}
\end{table}

We first note that all groups are captured well when $n >p$. In general, average number of mis-classifications remains very small ($<1$) for non-group variables, and so is the case of FP. The same is also true for overall FN when censoring is at low level, 20\%. We do observe, as in the earlier simulations,  FN increases with the increase in censoring, Cn, while falls with the increase in $n$. We further note that when censoring is low, both types of groups are captured well. As censoring increases, group capturing percentage drops for the continuous group.    

\section {Analysis of LIFE STUDY: Endocrine Disrupting Chemicals and Fecundity}
\label{LIFE}
\begin{figure}[t]
\centerline{\includegraphics[scale=0.9,width=15cm, height=10cm]{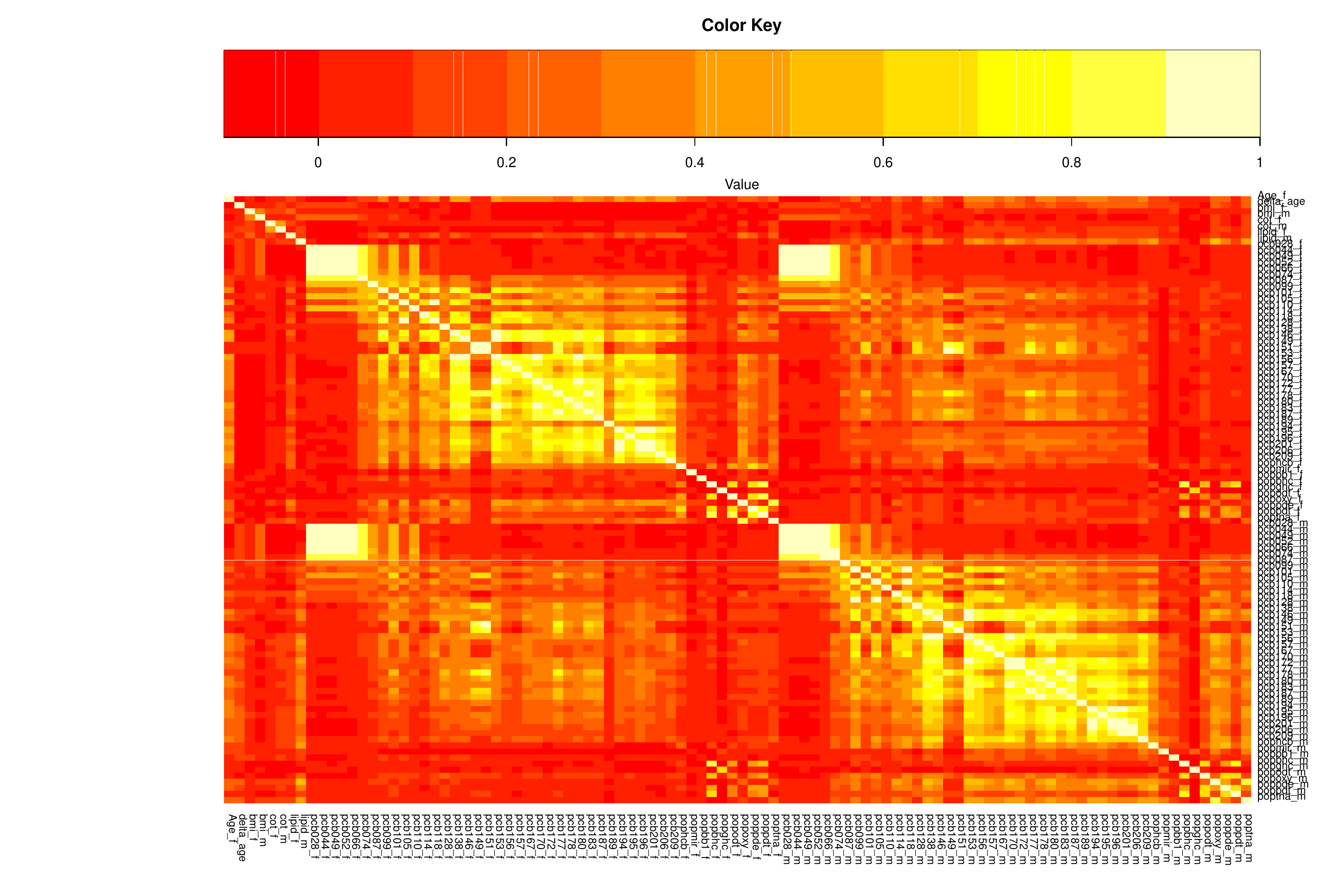}}
\caption{ Correlation heatmap among log-transformed variables. Yellow patches highlight zones of high positive correlation, and we see such all over, a bit blurry between groups (such OCP's, PCB' s etc) while quite prominent within a group. \label{100chems}} 
\end{figure}

In this section, we investigate the association of endocrine disrupting chemical concentration (in gm/cm$^3$) from both partners on time-to-pregnancy, which were recorded in a prospective cohort study, namely The Longitudinal Investigation of Fertility and the Environment Study (LIFE), conducted between 2005 and 2009. \citep{buck2011designing} The study targeted couples planing for pregnancy in next 6 months ($n= 501$), and they were monitored for next 12 months unless they get pregnant, or were lost to follow-up before the end of study. In addition, there could be cases when the monitoring process may have started late after couple of cycles ($max(L)= 3, L-1 \text{ cycles} $ truncated ). Thus,  time-to-pregnancy  is both right censored as well as left-truncated. The study recorded various chemical exposures in blood, among which our focus lies on Cotinine (indicative of smoking), lipid, and three classes of persistent organic pollutants (POP's), including  9 organochlorine pesticides (OCPs, such as $\beta$-hexachlorocyclohexane ($\beta$-HCH) \footnote{In the later sections, they are referred to by short-hand notations as follows;  OCPs and PBB are prefixed with \texttt{Pop-} while PCB's with \texttt{Pcb-}. For example, $\beta$-HCH measured in females are referred as \texttt{Popbhc$_f$}}, $p'$- dichlorodiphenyltrichloroethane (DDT) etc), polybrominated biphenyl (PBB-153), and 36 polychlorinated biphenyls (PCB's) in addition to important risk factors, female age, delta (difference between male and female partner's age) and bmi (in kg/m$^2$) for each partner of the couple. Our focus on these toxicants is motivated by  previous epidemiological studies, that indicate single or individual class of toxicants' exposure in pre-conception period may be adversely associated with fecundity, fetal growth and birth outcomes, such as birth weight, birth size etc. \citep{govarts2012birth, robledo2015preconception} However, studying this in the context of multi-pollutant chemical exposures, ie, mixtures of chemical toxicants on fecundity has not been conducted. This was partly due to lack of available statistical methods to address this question. Performing  a standard discrete survival analysis including all the toxicants poses few immediate challenges. As the number of variables, $p=100$, is quite large compared to sample size, and the variables are strongly correlated (see Figure \ref{100chems}), applying standard discrete frailty model may not produce efficient stable estimates, and thus necessitates a novel variables selection methodology, that addresses stable variable selection in discrete frailty modeling, in addition to right censoring and left truncation. In this context, elastic net penalization-based \texttt{Discnet} suits the purpose. This approach also helps identify the ``important drivers" of the mixtures of chemical toxicants.

Due to limited availability of biospecimens, chemical toxicants have missing values, which were imputed using standard methods, and 10 imputed copies of the data are analyzed separately to draw robust inferences. Since most of the chemicals register a low positive value, they are  highly skewed to 0.  For this reason, a log transform with offset 1\footnote{We note that an offset 1 leads to $\log(1+x) \approx x $ when x is small. }, $f(x) = \log(1+x)$, followed by standardization to have unit variance,  is used to lessen skewness and stabilize the scale of the estimates. Thus a full candidate model, subjected to variable selection, can be given by,
\begin{align*}
	g(\lambda_{it}) & = \gamma_{it}+ \beta_0+ \beta_1 \rm{Age_f +\beta_2 \,Delta\_age +\beta_3 \, Bmi_f+ \beta_4 \, Bmi_m+ \beta_5 \, Cotinine_f + \beta_6 \, Cotinine_m+ \beta_7 \, Lipid_f }  
	\\ & 
	 \rm{ + \beta_8 \, Lipid_m+ \beta_9 \, Pcb028_f+ \cdots + \beta_{100} \, Poptna_f+ b_i,}\,\, \rm{\quad  g(.)= logit, \quad \lambda_{it}}: \text{hazard}   
 \end{align*}	

In addition, there are three other challenges that require special attention- accommodating $-$ a) lipid variable and its binding effect with some lipophilic chemicals in the model, b) chemicals with high concentrations below limit of detection (LOD), and c)  potential non-linear effects of some chemicals.  We discuss in detail next.

\subsection{Accounting for lipophilic chemicals}
Some chemicals, such as POP's,  tend to bind with lipid in blood, thereby higher presence of lipid would induce higher amount of the chemicals present in the blood due to binding. They are known as lipophilic chemicals. To give an example,  abdelouahab et al (2013) \citep{abdelouahab2013maternal} showed  free thyroid hormone is associated with higher levels of polybrominated diphenyl ether (PBDE) when PBDE was expressed on a lipid basis, but not when PBDEs were expressed on a wet-weight basis.  In current literature on modeling lipids and lipophilic chemicals  often one of the two approaches is followed depending on whether a direct causal relationship can be attributed to lipid or not; -1) lipid as a covariate \citep{schisterman2005lipid}, -2) lipid as a concomitant \footnote{ A concomitant is a nonconfounding covariate which helps improve the precision of the estimate of interest when other variables are adjusted for it. As an example, height plays a role of concomitant variable when BMI is used as a surrogate for obesity instead of weight} variable \citep{li2013lipid}. 

While lipid as a covariate is straightforward, lipid as a concomitant requires new methods to be investigated when lipid is not assumed to be causally related . In the study of birth-defect, Li et al (2013) \citep{li2013lipid} suggested an adjustment to chemical-to-lipid ratio by replacing lipid with a boxcox transformation on it bringing in a power parameter, $\kappa$ in the modeling of a binarized birth outcome. They pointed out further generalization to the adjustment may be required in a given problem. One drawback of their adjustment is that a known function of lipid is only allowed to act as a factor, flexibly inflating or deflating the chemical. In view of this, we propose slightly generalized Box-Cox transformation to the chemical-to-lipid ratio while modeling infertility, a binary outcome variable, as a first stage of a two-stage procedure. In the second stage, disnet is performed on the standardized transformed variables based on estimated $\kappa$ with discrete survival outcome. Thus, if $s_i$ is the lipid level of $i^{th}$ individual and $x_i$ is the corresponding chemical concentration, we propose a two-stage procedure as follows, 
\begin{boxtext}
\section*{Two stage procedure for treating lipid as a concomitant}%
Stage 1: Fit  logit $(p_i) =  \alpha_0 + \alpha_1 x_i^*,  x_i^* = \text{standardized}   \, g(x_i,s_i; \kappa)$,  for each chemical using Bayesian logistic regression,
\begin{eqnarray*}
	&& g(x_i,s_i; \kappa)= \text{BxCx}(\frac{log(k_x+x_i)}{log(1+s_i)}, \kappa),\, log(k_x+x_i) >0, \\
	&& \text{BxCx}(y; \kappa)= \begin{cases}
		(y^\kappa -1)/\kappa,&  \kappa \ne 0 \\ \log(y), & \kappa = 0, 	
	\end{cases}  	
\end{eqnarray*}

\hspace{1cm} Estimate $\kappa$ by posterior median $\hat{\kappa}$ \\
Stage 2: 
Fit \texttt{Discnet} with the transformed variables, \text{standardized}   $\, g(x_i,s_i; \hat{\kappa})$ as defined in Stage 1
excluding lipid.    
\end{boxtext}

In above formulation $p_i$ denotes probability of being infertile, a binary outcome variable. A couple is treated infertile if they do not get pregnant in next 12 months. The fitting methodology only tweaks Li et al's procedure by defining a more generalized Box-Cox function. Every other aspect of MCMC computation implementing fast logistic regression remains the same. For the sake of completion, we outlined the MCMC steps in Web Appendix D in the supporting information.    

\subsection{Issue of limit of detection (LOD), and suspected non-linear effects }
Since most of chemicals are often found in blood in small proportion, their measurement suffers from what is known as limit (LOD) of detection issue. In other words, any registered value below a threshold, called limit of detection, may not be reliable. Though quite many approaches were suggested in the literature for bias and variance correction \citep{schisterman2004limit}  for such noisy data, two popular approaches to treat them, that we have implemented here, are:  1) replace all the values below LOD by $LOD/\sqrt{2}$, [referred to as ``With LOD treatment" or  WL], 2) keep all machine reported values as it is [referred to as WOL]. 

\begin{figure}[t]
	\includegraphics[scale=0.9,width=15cm, height=6cm]{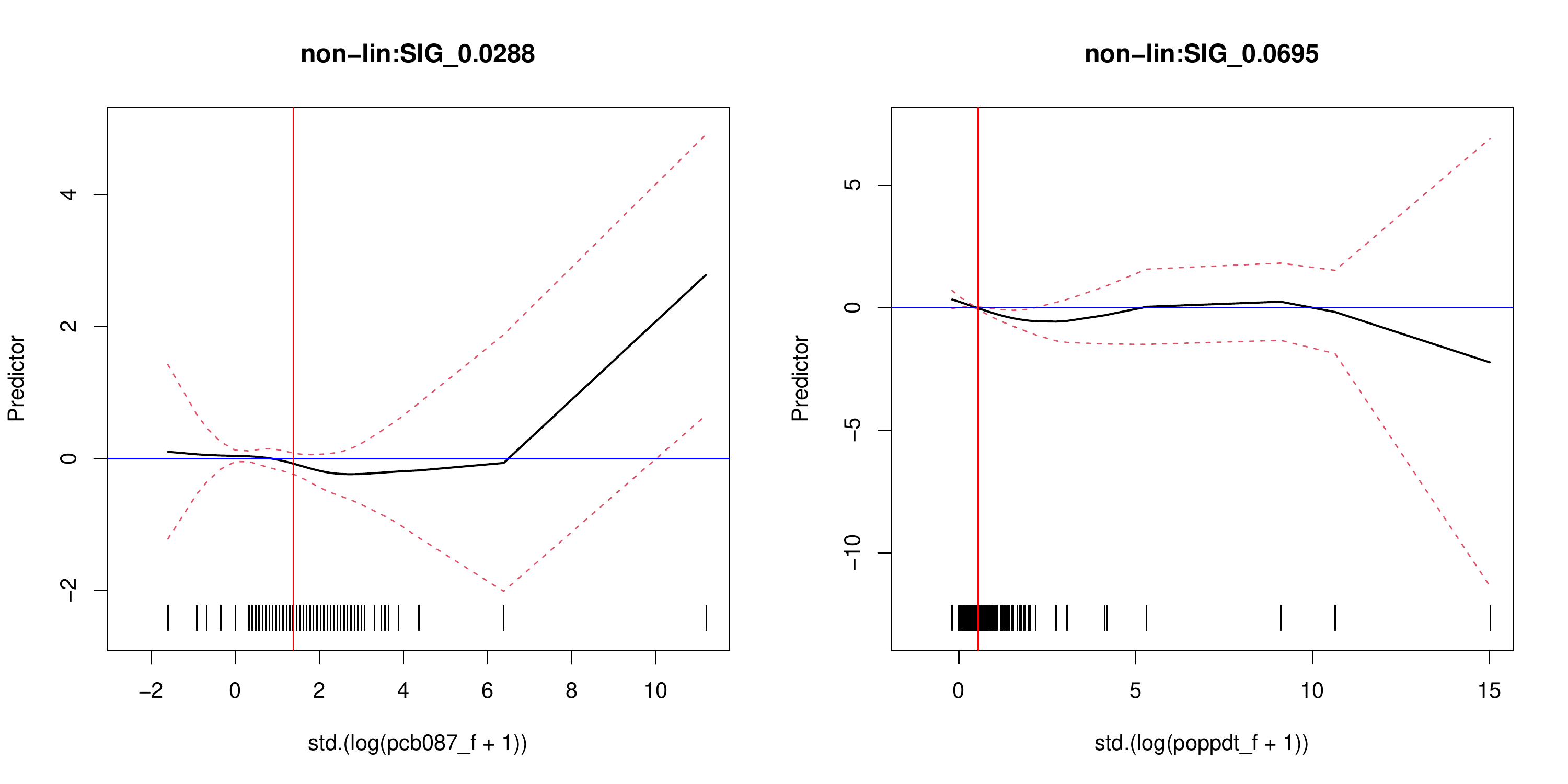}
	\caption{The attached \texttt{termplots} represent chemicals, that when kept in an unadjusted model, show significant non-linear effect ( p-value  $ < 0.1 $ ) in explaining  time to pregnancy. Each of them is converted to a binary variable, with 1 indicating values crossing a threshold, above which non-linear effect is experienced. \label{2chems} }
\end{figure}

Apart from LOD, one of the drawbacks of our framework is the assumption of linear effects of covariates in log-odds ratio. In this context, some chemicals, long suspected of non-linearity, need to be dealt with differently. To address this we again adopt two separate approaches, 1) without any non-linearity treatment, 2) Binarizing potential non-linear covariate. To elaborate the second approach, we first fit penalized spline with each chemical in an unadjusted model, then identify chemicals based on how severe (significant) the estimated non-linearity is. We have used \texttt{coxph} function from \texttt{survival} package along with pspline function for non-linearity. We then converted them to binary variable based on a threshold, above which non-linearity is suspected. 22 such chemicals are found based on 10\% significance level (See Web Appendix E for more details). Figure \ref{2chems}  represents two such chemicals, Pcb087$_f$, DDT$_f$ , for whom \texttt{termplot}s are shown. The red vertical line dictates approximately the point above which non-linearity is perceived. These 22 variables are replaced by their binarized versions in the full model.

\subsection{Variable selection and post-selection inference} \label{result2}
So far, we have considered 8 scenario's (Non-linearity ($\times 2$), Lipid treatment ($\times 2$), LOD ($\times 2$)). Besides these, we further screen out chemicals which have high proportion of values below LOD and re-ran the 8 scenatrio's for a smaller subset of 47 variables. The results from these analyses are contrasted with that of the full set of chemicals in the Web Appendix F. \texttt{Discnet} methodology is applied to the model with transformed variables, possibly preceded by a fitting procedure such as generalized Box-Cox model. \texttt{Discnet}  is composed of two steps: \emph{Step 1}. Finalizing variable selection, \emph{Step 2.} Fit the model without penalizing chemicals, while penalizing baseline for robust estimation.  In \emph{Step 1} Since there are 10 imputed datasets, \texttt{Discnet} is performed on each of them, and set of chemicals  selected in at least 6 of 10, are considered for refitting final model. \texttt{Discnet.R}  can be implemented after specifying a set of grid points for tuning parameters. We have used same set of grid points, and tuning mechanism for $\nu$, $\nu_s$, $\alpha $, as used  in simulation settings (See Section \ref{simul}).    

In \emph{Step 2} we re-fitted the same model with the selected variables only, with no further penalization on them (except on baselines). We obtain the parameter estimates using iterated weighted least square, and estimated asymptotic variances for each of the imputed data. Eventually parameter estimates and their standard deviations from 10 imputations are summarized in  a single table using \emph{Rubin}'s rule. Table \ref{tab4} gives odds ratio and its 95\% CI's for scenarios prior to non-linearity treatment.

\begin{table}[h]
	\centering
	\caption{Post selection inference: selected variables and estimated OR across various modes of analysis. Each blank cell indicates non-selection of that chemical in Step 1. \label{tab4}}
	\begin{tabular}{lcccccccc}
		\toprule
		\multicolumn{5}{c}{Linear effects assumed for chemicals}  \\\hline
		\multicolumn{3}{c}{Lipid(covariate)} & \multicolumn{2}{c}{ Lipid(concomitant)}   \\
		\cline{1-3}  \cline{4-5}
		Chems    & WOL   & WL  & WOL    & WL     \\\hline 
		(Intercept) & 0.15 (0.12, 0.17) & 0.15 (0.12, 0.17) & 0.14 (0.12, 0.18) & 0.14 (0.12, 0.18)\\
		Cotinine$_f$ & 0.82 (0.7, 0.95) & 0.79 (0.68, 0.91) & 0.8 (0.69, 0.92) & 0.84 (0.72, 0.98)\\
		Cotinine$_m$ & 0.9 (0.79, 1.03) &  &  & 0.89 (0.77, 1.01)\\
		Age$_f$ & 0.8 (0.71, 0.9) & 0.81 (0.72, 0.91) & 0.82 (0.73, 0.93) & 0.83 (0.73, 0.93)\\
		Pcb049$_f$ & 1.05 (0.78, 1.42) & 1.13 (0.59, 2.18) &  & \\
		Pcb028$_f$ &  &  & 1.14 (1.02, 1.28) & 1.15 (1.02, 1.29)\\
		Pcb028$_m$ &  & 1.07 (0.54, 2.1) &  & \\
		Pcb052$_m$ & 1.04 (0.74, 1.45) &  &  & \\
		Pcb101$_m$ & 1.13 (1, 1.28) &  &  & \\
		Pcb044$_f$ &  & 1.11 (0.56, 2.23) &  & \\
		\bottomrule
	\end{tabular}
\end{table}
We note the female smoking behavior (cotinine levels) and female age are negatively associated with fecundity, and thus increase risk across all 4 modes of analysis, while male's smoking behavior seems equally important as it seems to be associated in two sub-scenarios.  Although quite a many PCB's  got selected across various modes of analysis, they may not be strongly associated with pregnancy risk. Among them, Pcb028$_f$ seems to be positively associated with fecundity when treated as a function to ratio-to-lipid. Pcb028$_m$ turns out to be positively associated when lipid is treated as a covariate, and no LOD treatment is done. Table \ref{tab5}, on the other hand lists similar figures when chemicals suspected of non-linearity are converted to binary variables. 

\begin{table}[h]
	\centering
	\caption{Post selection inference: chemicals, suspected of non-linear effect, are binarized based on a threshold, above which non-linearity is suspected. Chemicals asterisked (*) represent the set of such cases, as selected by \texttt{Discnet}. Among them, only DDT$^*$ seem to be significantly negatively associated with the pregnancy across most of the cases. Chemicals marked $\dagger$, on the  other hand, represent cases having at least 50\% of values above LOD.\label{tab5}} 
	\begin{tabular}{lcccccccc}
		\toprule
		\multicolumn{5}{c}{ Suspected non-linear effects accommodated}  \\\hline
		\multicolumn{3}{c}{Lipid (covariate)} & \multicolumn{2}{c}{ Lipid (concomitant)}   \\
		\cline{1-3}  \cline{4-5} 
		Chems    & WOL   & WL  & WOL    & WL     \\\hline      
		(Intercept) & 0.17 (0.14, 0.2) & 0.16 (0.13, 0.19) & 0.15 (0.12, 0.18) & 0.16 (0.13, 0.19)\\
		Cotinine$_f$ & 0.82 (0.7, 0.95) & 0.82 (0.7, 0.96) & 0.8 (0.69, 0.92) & 0.83 (0.71, 0.97)\\
		Cotinine$_m$ & 0.91 (0.79, 1.04) & 0.89 (0.77, 1.02) &  & 0.88 (0.77, 1.01)\\
		Pcb194$_m^{*,\dagger}$ & 0.85 (0.64, 1.12) &  &  & \\
		Age$_f$ & 0.86 (0.75, 0.98) & 0.85 (0.75, 0.96) & 0.82 (0.73, 0.93) & 0.85 (0.75, 0.97)\\
		DDT$_f^*$ & 0.75 (0.57, 0.98) & 0.75 (0.57, 0.98) &  & 0.78 (0.59, 1.03)\\
		Pcb049$_f$ & 1.06 (0.77, 1.47) & 1.17 (0.61, 2.25) &  & \\
		Pcb028$_f$ &  &  & 1.14 (1.01, 1.28) & 1.13 (1.01, 1.27)\\
		Pcb028$_m$ & 1.02 (0.54, 1.9) &  &  & \\
		Pcb052$_m$ & 1.08 (0.69, 1.69) &  &  & \\
		Pcb044$_f$ &  & 1.13 (0.58, 2.19) &  & \\
		Pcb101$_m$ & 1.13 (1, 1.29) &  &  & \\
		\bottomrule
	\end{tabular}
\end{table}
We note that in this case, \texttt{Discnet} retains all variables as were selected in Table \ref{tab4}. We also observe odds ratio and their 95\% CI levels  remain almost the same.  When compared with Table \ref{tab5}, Table \ref{tab4} fails to pick any chemical from the list, suspected of non-linearity. On the other hand Table \ref{tab5} selected two such chemicals, namely Pcb194$_m$, DDT$_f$, one of which is well known harmful chemical, DDT$_f$, which came negatively associated with TTP across all 4 modes of analysis. Thus, \texttt{Discnet} is capable of selecting important chemicals while making robust inferences about them.   
\section{Discussion}
This article proposes a variable selection approach to identify the important dirivers in the mixtures of endocrine disrupting chemicals on fecundity. In our analysis of  the LIFE Study, we identified Female Cotinine and Female DDT as important drivers of the mixtures, with both of them negatively associated with fecundity. These findings were upheld over various modes of analysis. To the best of our knowledge, this is the first method that is applicable to assess mixture of chemical toxicants association to time-to-pregnancy, a discrete survival time subject to left truncation and right censoring. 

We have proposed a novel variable selection, and fitting methodology, \texttt{Discnet}, in discrete hazard model with frailty via elastic net penalty for strongly correlated covariates, where survival time is both right censored and left-truncated. The elastic net-based method proposes a tuning mechanism that chooses $\lambda$ (lasso-part) penalty using permutation method, and while others using BIC. \texttt{Discnet} improves the performance compared to its peers in presence of correlated covariates both in terms of stability and FP rates. It seems to work very well, in either $n >p$ scenario and/or  low censoring cases, otherwise it performs at par with other methods.  In the time to pregnancy study, we performed variable selection, and subsequent model fitting in multiple scenarios corresponding to multiple ways to deal with limit of detection issue, treatment of lipid in presence of lipophilic chemicals, as well as for the treatment of non-linear effects. In the process, we have suggested a new method based on generalized box-cox transformation  that treats lipid as a concomitant variable. 

There are several aspects which have room for improvements for future extensions. The hazard model can be easily extended to cloglog model.
We are currently working to incorporate interactions between covariates with heredity constraints  for discrete hazard model with frailty while performing variable selection. It would be interesting to see how various choices of loss functions impact the variable selection strategy. We plan to incorporate nonlinear effect of covariates in  the variable selection strategy through a generalized additive model type frameworks.

In this paper, we have focused on the frequentist approaches for variable selection, which among other desirable properties are easier to compute. Alternatively,  computationally intensive Bayesian non-paramteric methods such as Bayesian kernel machine regression with variable selection (BKMR) have been shown to perform better than L1-penalized methods \citep{lazarevic2019performance, bobb2015bayesian}, especially in the context of chemical mixtures with continuous outcomes. However. they need to be investigated more thoroughly in the context of survival outcomes subject to left truncation and right censoring.



\section*{Acknowledgments}
This research was supported by the Intramural Research Program of the {\it Eunice Kennedy Shriver} National Institute of Child Health and Human Development [contracts N01-HD-3-3355, N01-HD-3-3356, and N01-HD-3-3358].  The authors would like to acknowledge that this study utilized the high-performance computational capabilities of the Biowulf Linux cluster at the National Institutes of Health, Bethesda, MD (\texttt{http://biowulf.nih.gov}).

\subsection*{Financial disclosure}

None reported.

\subsection*{Conflict of interest}

The authors declare no potential conflict of interests.

\section*{Supporting information}

Web Appendices, Tables and Figures, referenced in Sections \ref{simul}, \ref{LIFE},  are submitted in \texttt{Supporting\_information.pdf}. \texttt{Discnet} algorithm is implemented in an R-function by the same name in \texttt{Discnet.R} with a host of auxiliary subroutines supporting it. They can be provided if required.  



\vspace{-0.25 cm}
\appendix
\section{ Treatment of a concomitant variable in a survival model \label{app1}}
A concomitant variable is defined to be a variable that may not directly influence an outcome but may influence the relationship of a covariate with the outcome variable. For example, Li et al \citep{li2013lipid} treated lipid as a concomitant  variable which may influence the relationship between lipophilic chemicals and a health outcome of interest. In the context of survival model, the link function on discrete hazard is modeled as, 
\vspace{-0.7cm}     
\begin{equation*}
	\eta_{i,t}= \gamma_t + \beta_0 + \sum_j^p \beta_j* f_j(x_j,s) + b_i,
\end{equation*}

where $s$ is treated as a concomitant variable which impacts the relationship of covariate $x_j$'s with the TTP--captured by the function  $f_j$. Li et al used $f_j(x_j,s)= x_j/g(s; \kappa_j), g(s; \kappa_j)$ = $Boxcox(s,\kappa_j)$ in their study with lipids, and estimated  $\kappa_j$'s using a Bayesian logistic model (note that a binary health outcome was modeled in their study). In Section 5, while treating lipid as a concomitant we worked with slightly more generalized BoxCox function (as defined in Section 5) for $f_j(x_j,s)$ and implemented Bayesian logistic model treating infertility as health outcome to obtain  estimates for $\kappa_j$'s to be used in second stage analysis of variable selection and association study.
\vspace{-0.5cm}
\section{Modified gradient ascent with EM type updates \label{app2}}

\begin{algorithm}
	\caption{Coordinate-wise gradient (modified) ascent}
	\label{Myalgo}
	For a fixed values of the tuning parameters, $\nu, \alpha$ and  $\nu_s$,\\
	\textbullet ~\textbf{Definitions}: 	
	\begin{algorithmic}[1]
		\State  $ \bss{\theta} $: Set of all parameters except  $\bs{Q}$  
		\Statex $\theta_0 = \beta_0;\quad \theta_i =\beta_i \quad \forall  i=1,\cdots,p; \quad \theta_{(p+1+j)} = \gamma_j, \quad \forall j=1,\cdots,t_{max}$  
		\Statex	$\theta_{(p+1+t_{max}+k)} = b_k \quad \forall k=1,\cdots,qn$		 
		\State  $\bs{H}^T: (\bs{X}^T,\bs{A}^T,\bs{Z}^T);$
		\State $\bss{\lambda}(\bss{\theta})^T= (\lambda(s=1|\bs{x}_1;\bss{\theta}),\cdots,\lambda(s=t_1|\bs{x}_1;\bss{\theta}),\lambda(s=1|\bs{x}_2;\bss{\theta}),\cdots,\lambda(s=t_n|\bs{x}_n;\bss{\theta}))^T$.
	\end{algorithmic}
	
	\textbullet \textbf{Step1}: Initialize parameters of interest ,$(\bss{\theta}=\bss{\theta}^{(0)} =(\beta_0^{(0)},\bss{\beta}^{(0)},\bss{\gamma}^{(0)},\bs{b}^{(0)}), \bs{Q}=\bs{Q}^{(0)})$\\
	Iterate over $l=1,2,\ldots,$ until convergence \\
	\textbullet \textbf{Step2}:  Block-update $\bss{\theta}=\hat{\bss{\theta}}^{(l)} | \hat{\bss{\theta}}^{(l-1)}, \hat{\bs{Q}}^{(l-1)}$ as follows,  
	\begin{algorithmic}[1]  
		\State \emph{Compute modified gradient of penalized likelihood}, $S^{pen}(\bss{\theta})$ at $\bss{\theta}^{(l-1)}$, 
		\Statex With $ S(\bss{\theta} ) =  \partial l(\bss{\theta})/\partial \bss{\theta} =  \bs{H}[\bs{y}-\bs{\lambda(\bss{\theta})}]$,
		\begin{eqnarray*}
			S^{pen}_0(\bss{\hat{\theta}}^{(l-1)}) &=& S_0(\bss{\hat{\theta}}^{(l-1)}),  	\\	
			\begin{aligned} &S^{pen}_i(\bss{\hat{\theta}}^{(l-1)}) , \\ 
				& i = 1,\cdots, p, 
			\end{aligned}
			&=&   \begin{cases} 
				S_i(\bss{\hat{\theta}}^{(l-1)})- \nu(1-\alpha)\hat{\beta}_i^{(l-1)} - \nu \alpha * sign(\hat{\beta}_i^{(l-1)})   \text{, if } \hat{\beta}_i^{(l-1)} \ne 0 \\
				S_i(\bss{\hat{\theta}}^{(l-1)}) - \nu \alpha * sign(S_i(\bss{\hat{\theta}}^{(l-1)}))  \text{, if } \hat{\beta}_i^{(l-1)} = 0 \,\,\text{and}\,\, |S_i(\bss{\hat{\theta}}^{(l-1)})| \\
				\quad \quad \quad \quad> \nu\alpha \\  
				0   \text{, Otherwise} \\                                                                          
			\end{cases}  \\
			S^{pen}_i(\bss{\hat{\theta}}^{(l-1)}) &=&  
			S_i(\bss{\hat{\theta}}^{(l-1)})- \nu_s\hat{\gamma}_{j}^{l-1},\quad  i = p+1+j; j=1,\cdots,t_{max},\\
			S^{pen}_{\mathcal{I}}(\bss{\hat{\theta}}^{(l-1)}) &=&  
			S_{\mathcal{I}}(\bss{\hat{\theta}}^{(l-1)})- \bs{Q}_b^{-1}\bs{b},\quad \mathcal{I} = (p+1+t_{max}+1, \cdots,p+1+t_{max}+qn)^T,\\
			sign(x) &=&  I(x > 0) - I(x < 0) 		 
		\end{eqnarray*}	 	
		\State \emph{Compute the directional second derivative at any $\bss{\theta}$ and at any direction $\bs{v}$ },
		$l_{pen}''(\bss{\theta},\bs{v})= -\bs{v}^T  F^{pen}(\bss{\theta}) \bs{v} $, where  $F^{pen}(\bss{\theta}) =-E(\nabla^2 l_{pen}(\bss{\theta})$= Fisher's matrix. Note,
		\begin{eqnarray*}
			F^{pen}(\bss{\theta})	&=&  \bs{H}\bs{W}\bs{H}^T + \bs{K}, \,\, \bs{K} = Diag(0,\nu(1-\alpha)\bs{I},\nu_s\bs{I}, \bs{Q}_b^{-1} ),\\
			\bs{W}&=& Diag(.,h(\eta_{ij})(1-h(\eta_{ij})),.) \,\, \text{wherever $S^{pen}$ is differentiable};  
		\end{eqnarray*}
		Note further, $F^{pen}(\bss{\theta})$ can be extended by continuity to all $\bss{\theta}$.
		\State \emph{Step size for gradient ascent step: compute $t_{opt}$ and $t_{min}$ as in Goeman (2010)\citep{goeman2010L1}}
		\begin{eqnarray*}
			t_{edge}^{(l-1)}&=& min_i \Big\{ - \frac{{\hat{\theta_i}}^{(l-1)}}{S_i^{pen}(\hat{\bss{\theta}}^{(l-1)})}: sign(\hat{\theta}_i^{(l-1)})= -sign(S_i^{pen}(\hat{\bss{\theta}}^{(l-1)})) \ne  0  \Big\}, \\
			\text{and } t_{opt}^{(l-1)}	&= &   -\frac{||S_i^{pen}(\hat{\bss{\theta}}^{(l-1)})||_2}{l''_{pen}(\hat{\bss{\theta}}^{(l-1)}, S_i^{pen}(\hat{\bss{\theta}}^{(l-1)}) )}, \,\, \text{ $||.||_2$ being	$L_2$ norm} \\
		\end{eqnarray*}
		\State \emph{Update $\bss{\theta} \mapsto \hat{\bss{\theta}}^{(l)}$  by choosing optimal step without changing signs in any coordinate}
		\begin{eqnarray*}
			{\hat{\bss{\theta}}}^{(l)} &=& \begin{cases} 
				{\hat{\bss{\theta}}}^{(l-1)} + t_{edge}^{(l-1)} S^{pen}(\hat{\bss{\theta}}^{(l-1)}) \quad \text{if }   t_{opt}^{(l-1)} \ge t_{edge}^{(l-1)}, \\
				{\hat{\bss{\theta}}}_{FS}^{(l)}                                                 \quad \text{if }   t_{opt}^{(l-1)} < t_{edge}^{(l-1)} \text{, and } sign({\hat{\bss{\theta}_{FS}}}^{(l)})=sign({\hat{\bss{\theta}}}_{+}^{(l-1)}),\\
				{\hat{\bss{\theta}}}^{(l-1)} + t_{opt}^{(l-1)} S^{pen}(\hat{\bss{\theta}}^{(l-1)})  \quad \text{otherwise}.
			\end{cases} \\
			sign({\hat{\bss{\theta}}}_{+}^{(l-1)}) &=& \lim_{\epsilon \downarrow 0}  sign(\hat{\bss{\theta}}^{(l-1)}+\epsilon S^{pen}(\hat{\bss{\theta}}^{(l-1)})), 
			{\hat{\bss{\theta}}}_{FS}^{(l)} =     {\hat{\bss{\theta}}}^{(l-1)} +  [F^{pen}(\bss{\theta})]^{-1}  S^{pen}(\hat{\bss{\theta}}^{(l-1)}),
		\end{eqnarray*}
	\end{algorithmic}
	\textbullet \textbf{Step3}:  Block-update $\bs{Q} \mapsto\hat{\bs{Q}}^{(l)} | \hat{\bss{\theta}}^{(l)}, \hat{\bs{Q}}^{(l-1)}$ as follows,  
	using EM-based update strategy \citep{groll2014variable},
	\vspace{-0.35cm}
	\begin{eqnarray}
		\hat{\bs{Q}}^{(l)} &=& \sum_{i=1}^{n} (\bs{V}_{ii} + \bs{b}_i^{(l)}{\bs{b}^{(l)}_i}^T)/n, \,\,  
		\bs{V}_{ii} = F_{ii}^{-1} + F_{ii}^{-1}F_{i \tilde{\bss{\theta}}} ( F_{\tilde{\bss{\theta}}\tilde{\bss{\theta}}} - \sum_{i=1}^n F_{\tilde{\bss{\theta}} i} F_{ii}^{-1} F_{i\tilde{\bss{\theta}}})^{-1}F_{\tilde{\bss{\theta}} i}F_{ii}^{-1} \nonumber
	\end{eqnarray}
	$F_{xy}$ is a submatrix of $F^{pen}$ that corresponds to rows that represent x and columns that represent y. In this case $\tilde{\bss{\theta}}^T= (\gamma^T, \beta_0 ,\bss{\beta}_{active}^T)$  (See Web Appendix A for more details)
\end{algorithm}



\newpage
\bibliography{wileyNJD-AMA}

\clearpage



\end{document}